\documentclass[aps,prl,showpacs,english,10pt,a4,nofootinbib,notitlepage,twocolumn,superscriptaddress]{revtex4-1}
\usepackage{dcolumn}    
\usepackage{ifpdf}
\usepackage{amssymb,lineno,amsfonts}
\usepackage{graphicx}   
\usepackage{dcolumn}    
\usepackage{bm}         
\usepackage{bbm}
\usepackage{mathrsfs}
\usepackage{upgreek}
\usepackage{mathtools}
\usepackage{epstopdf}
\usepackage{setspace}
\usepackage{hyperref}
\usepackage{natbib}
\usepackage[usenames,dvipsnames,table]{xcolor}
\usepackage[normalem]{ulem}
\usepackage{multirow,booktabs}
\graphicspath{{figures/}}

\usepackage{times}

\definecolor{med-blue}{RGB}{25,25,112}
\hypersetup{colorlinks, linkcolor={red},citecolor={blue}, urlcolor={MidnightBlue}}

\newcommand{\ket}[1]{\vert{#1}\rangle}
\newcommand{\bra}[1]{\langle{#1}\vert}
\newcommand{\outpr}[2]{\vert{#1}\rangle\langle{#2}\vert}

\definecolor{med-blue}{RGB}{25,25,112}
\hypersetup{colorlinks, linkcolor={red},citecolor={blue}, urlcolor={MidnightBlue}}

\usepackage{color}
\definecolor{dred}{rgb}{.8,0.2,.2}
\definecolor{ddred}{rgb}{.8,0.5,.5}
\definecolor{dblue}{rgb}{.2,0.2,.8}
\definecolor{dgreen}{rgb}{.2,0.5,.2}

\newcommand{\be}{\begin{equation}}
\newcommand{\ee}{\end{equation}}

\newcommand{\bea}{\begin{eqnarray}}
\newcommand{\eea}{\end{eqnarray}}

\begin{document}

\title{Experimental violation of the Leggett-Garg inequality in a 3-level system}
\author{Hemant Katiyar}
\email{hkatiyar@uwaterloo.ca}
\affiliation{Institute for Quantum Computing, University of Waterloo, Waterloo N2L 3G1, Ontario, Canada}
\affiliation{Department of Physics and Astronomy, University of Waterloo, Waterloo, Ontario N2L 3G1, Canada}
\author{Aharon Brodutch}
\email{brodutch@physics.utoronto.ca}
\affiliation{Institute for Quantum Computing, University of Waterloo, Waterloo N2L 3G1, Ontario, Canada}
\affiliation{Department of Physics and Astronomy, University of Waterloo, Waterloo, Ontario N2L 3G1, Canada}
\author{Dawei Lu}
\email{d29lu@uwaterloo.ca}
\affiliation{Institute for Quantum Computing, University of Waterloo, Waterloo N2L 3G1, Ontario, Canada}
\affiliation{Department of Physics and Astronomy, University of Waterloo, Waterloo, Ontario N2L 3G1, Canada}
\author{Raymond Laflamme}
\email{laflamme@uwaterloo.ca }
\affiliation{Institute for Quantum Computing, University of Waterloo, Waterloo N2L 3G1, Ontario, Canada}
\affiliation{Department of Physics and Astronomy, University of Waterloo, Waterloo, Ontario N2L 3G1, Canada}
\affiliation{Perimeter Institute for Theoretical Physics, Waterloo N2L 2Y5, Ontario, Canada}
\affiliation{Canadian Institute for Advanced Research, Toronto, Ontario, Canada}

\begin{abstract}
The Leggett-Garg (LG) test of macroscopic  realism involves a series of   dichotomic non-invasive measurements that are used   to calculate a function which has a  fixed upper bound for a macrorealistic  system and a larger upper bound for a quantum system. The quantum upper bound depends on both the details of the measurement and the dimension of the system. Here we present an LG experiment on a three-level quantum system, which produces a  larger theoretical quantum upper bound  than that of  a two-level quantum system. The experiment is carried out in nuclear magnetic resonance (NMR) and consists of the LG test as well as a test of the ideal assumptions associated with the experiment, such as  measurement non-invasiveness. The non-invasive measurements are performed via the modified ideal negative result measurement scheme on a three-level system.  Once these assumptions are tested, the violation becomes small, despite the fact that the LG value itself is large.  Our results showcase the advantages of using the modified measurement scheme that can reach the higher LG values, as they give more room for hypothetical malicious errors  in a real experiment

\end{abstract}

\maketitle

\textit{Introduction.}---The predictions of quantum mechanics regarding microscopic systems do not carry over to macroscopic objects. Unlike photons and electrons, cats and tables do not seem to exist  in a superpositions of two classically observable states such as dead and alive or here and there. There is, however, no known theoretical limit on the size of objects that can be observed in an arbitrary  superposition of two states and it is conceivable that we will one day be able to isolate large objects from the environment, such that they can be in what we may call a macroscopic superposition state.  The Leggett-Garg (LG) experiment \cite{LG} and some extensions \cite{Emary2013,Knee2016,*Mao} allow us to test the assumption that a given system confined to a discrete set of \emph{classically observable} states is never in a superposition of these states. The experiment leads to an inequality that, under some reasonable assumptions, cannot be violated  when the system is in a definite classically observable  state at all times, but can be violated when it is superposition of these states.

Unlike Bell's inequality, the assumptions regarding the Leggett-Garg inequality (LGI) depend on the physical system and the experimental setup. Of the  three fundamental assumptions: (A1) macroscopic realism (MR): \emph{the system cannot be in a superposition of the classically observable state}, (A2) non-invasive measurability (NIM): \emph{It is possible to measure the macroscopic system without disturbing it}, and (A3) induction: \emph{the future cannot influence the past}, only the last is independent of the experimental setup. The LGI is therefor a test of  MR under a set of reasonable assumptions about the system, in particular a version of NIM. The violation of the inequality leads to the conclusion that either MR or one of the other assumptions is incorrect \cite{Maroney2014}. The aim of a well-designed experiment is therefore to convince a skeptic that the incorrect assumption is probably  MR, i.e. the system is in a superposition of classically observable states sometime during its evolution.

Various experimental tests of 2-level LGI have been performed \cite{Palacios,opto,deco,photons,semiweak, maheshnmr,souza,e-trans,back,quantumdot,crystal,knee2012violation,katiyar2013violation}, and in \cite{knee2012violation,katiyar2013violation} ideal negative result measurements (INRMs) were used to perform non-invasive measurements. However, none of these previous experiments tested the assumption that the measurements are non-invasive. A slight modification of the experiment was recently introduced in \cite{Knee2016}, as far as we know that is the first experiment where the NIM assumption was tested.

In the  standard  LG experiment a parameter $K_3$ is classically constrained to take values between $-3$ and $1$, whereas quantum mechanics predicts possible violations of up to $1.5$, giving a narrow margin for experimental errors. It has recently been noted  \cite{Knee2016,*Mao} that in order to convince a skeptic that the reasonable assumptions  are indeed reasonable, they need to be tested and the inequality must be adjusted accordingly. Consequently the margin for error gets reduced even further. Until recently it was  believed  that the maximal violation of $K_3$    is independent of the number of possible macroscopically distinct states of the system due to the fact that the measurements are dichotomic.  However,  Budroni  and Emary \cite{emary_base}  showed  that this is only true if the measurements follow the naive L\"uders update rule. In a more general setting, it is possible to observe larger violations by going to higher dimensional systems, up to the algebraic maximum of 3.  While such measurements give a bigger margin for errors, it was not clear how to construct them in a reasonable way that does not require a seemingly  artificial dephasing step  between measurements which is in conflict with NIM.

In this work we  demonstrate the first  violation of the LGI with an experiment that has a theoretical bound beyond $K_3=1.5$.  We present results of  a set of experiments performed on an  ensemble of  3-level systems in liquid-state nuclear magnetic resonance (NMR)  and  provide a natural method for performing the required measurement without an artificial dephasing step. The inequality is corrected for  a number of non trivial  assumptions about the state of the systems and the measurement device, in particular the LGI is corrected to account for non   ideal  measurements.

\begin{figure}[htb]
\begin{center}
	\includegraphics[width=0.95\columnwidth]{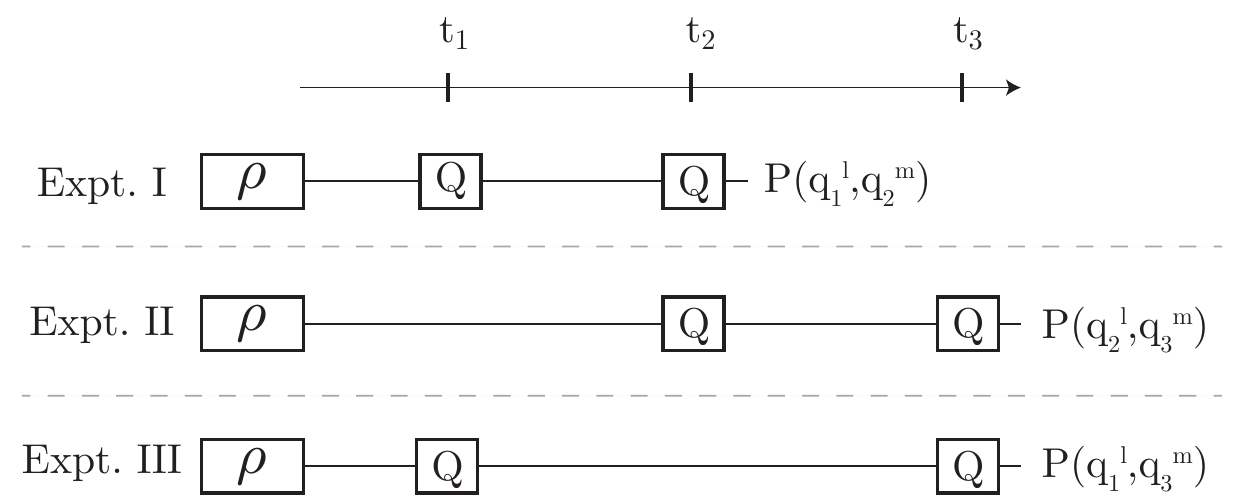}
\end{center}
	\caption{Scheme for LG test. Three experiments are performed  such that  in each experiment  the dichotomic  observable $ Q $ is measured at two different times.  The experiments are  performed many number of times to obtain the probabilities for the correlations and  anti-correlations between two measurements. The results are used to calculate the LG string in Eq. (\ref{eq:k3}). }
	\label{fig_lgi}
\end{figure}

\textit{The Leggett-Garg test.}---Consider a system which is evolving under certain Hamiltonian.
We decide to perform  dichotomic  measurements of an observable $Q$, at some time $ t_i $ represented as $ Q_i $, that can perfectly distinguish between two states  of a system.
The outcomes of these measurements  are denoted
as $ q_i^1=+1$ and $ q_i^2=-1$.
{In a macrorealistic system, the outcomes $q_i^l$ ($l=1, 2$) represent the \emph{real} state of the system, i.e. if the result was $q_i^l$ we can infer that the system was in the state corresponding to $q_i^l$ at time $t_i$. A test of macrorealism is a test of this hypothesis.}
For LG test, one chooses three distinct times to perform a measurement and three independent experiments.
In each of the three experiments we start with the same state, and  then perform measurements on two of the three chosen times as shown in Fig. \ref{fig_lgi}.
These three independent experiments are performed many times to estimate the probabilities of being in different possible states.
Using these probabilities one can calculate the two time correlations of the measurements,
\begin{equation}
\langle Q_i Q_j \rangle = \sum\limits_{l,m}q_i^l q_j^m P(q_i^l,q_j^m),
\label{eq:correlation}
\end{equation}
where $q_i^l$ ($l=1, 2$) means the $l^{th}$ outcome of measurement performed at $t_i$.    

The 3-measurement LG string is
\begin{equation}
K_3 =  \langle Q_1 Q_2 \rangle + \langle Q_2 Q_3 \rangle - \langle Q_1 Q_3 \rangle.
\label{eq:k3}
\end{equation}

{If we assume that the measurements do not disturb the system (NIM assumption)} and the system is classical (i.e. macrorealistic), the value of $K_3$ is bounded by $-3\le K_3 \le 1$.
On the other hand, if the system is quantum it is possible to choose the evolution times between measurements in such a way that $K_3$ will go beyond 1, violating the LGI that $K_3\le 1$. The quantum bound for $K_3$ is 1.5 for a 2-level system \cite{Emary2013}. More general systems have the same bound if the measurements follow the L\"uders update rule which is natural for these types of experiments.   According to the L\"uders rule the dichotomic  measurement projects the state of the system into one of two orthogonal subspaces corresponding to the $\pm 1$ measurement results. While this projection is invasive when the system is quantum, it is theoretically non-invasive if we assume MR.   In performing the LG test, we must however consider the possible objection of a skeptic who may object to our assumption that the measurement indeed follows the L\"uders rule.  To counter such an argument, LG suggested that the experiment is carried out using ideal negative result measurements (INRMs). INRMs are implemented  by measuring if a system is in a given state (say that state that corresponds to  $q_i^1=+1$) and post-selecting on negative outcomes that allow us to infer the state of the system, e.g.  by finding that the system  is not in a  $q_i^1=+1$ state  we infer that it must be in a $q_i^2=-1$ state.

The original LG test considered only 2-level systems. Recently Budroni and Emary \cite{emary_base} showed that if one relaxes the assumption that the measurement follows the L\"uders update rule, and instead one allows a more general update rule which also destroys some of the phase information within the $\pm 1$ subspaces, then the quantum bound on $K_3$ could be extended to a value that depends on the dimension of the system, and goes asymptoticly to the algebraic maximum  of 3. For a 3-level system, such measurements can lead to the value $K_3=1.7566$, when the observable $Q=-\ket{0}\bra{0}+\ket{1}\bra{1}+\ket{2}\bra{2}$ and the measurement acts like a complete dephasing channel.  However, the channel  seems to be more invasive than  necessary and can raise questions about the validity of NIM. In such a case, it is hard to justify the violation of the LGI as a violation  of MR. However, as we show below, the channel can be implemented using INRMs.

\begin{figure}[htb]
\begin{center}
	\includegraphics[width=0.95\columnwidth]{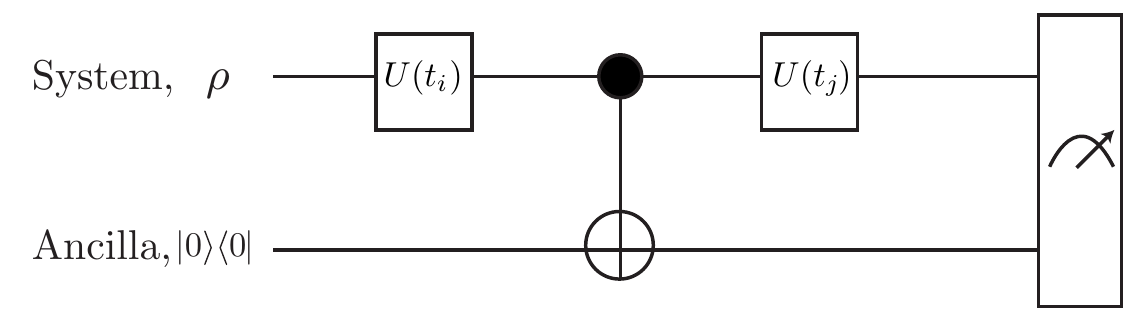}
\end{center}	
	\caption{General Scheme for a single run of the LG test with an INRM. We start with the system in some state $ \rho $ and ancilla in $ \outpr{0}{0} $.
		The two evolution times  $t_i$ and $t_j$ depend on  which of the three experiments is performed (see Fig. \ref{fig_lgi}). The controlled gate is the first measurement  performed (one of three possible  INRMs), and it is non-invasive if nothing happens, i.e the state of the ancilla is unchanged. The last measurement is not necessarily non-invasive since we are not concerned about the future dynamics of the system. The results are post-selected to include only the instances when the INRM was successful, i.e. the situations where the ancilla is in the state $ \outpr{0}{0} $. For each measurement setting in Fig. \ref{fig_lgi}, we perform three runs, one for each state of the system. }
		\label{fig_scheme1}
\end{figure}

\textit{Measuring the Probabilities using INRMs.}---The scheme for performing the modified LGI measurement is based on using three INRMs, one for each of the possible states. The measurement is registered on an ancillary qubit  initially in the state $\ket{0}$. When performing the INRM of the system state $\ket{j}$, the ancilla  remains in the state $\ket{0}$ if the system is in $\ket{j}$ and rotates to $\ket{1}$ otherwise. The three gates  below correspond to the three types of measurements.
\begin{eqnarray}
\textrm{CG$_0 $} = \outpr{0}{0}\otimes\mathbbm{1} + \outpr{1}{1}\otimes X + \outpr{2}{2}\otimes X, \\
\textrm{CG$_1 $} = \outpr{0}{0}\otimes X + \outpr{1}{1}\otimes \mathbbm{1} + \outpr{2}{2}\otimes X, \\
\textrm{CG$_2 $} = \outpr{0}{0}\otimes X + \outpr{1}{1}\otimes X + \outpr{2}{2}\otimes \mathbbm{1}.
\end{eqnarray}
Consider, for example, the application of CG$ _0 $ on the following general state of system and ancilla being in state $ \ket{0} $
\[\begin{bmatrix}
P_0 & a & b\\
a^\dagger & P_1 & c \\
b^\dagger & c^\dagger & P_2
\end{bmatrix}_S
\otimes
\begin{bmatrix}
1 & 0 \\
0 & 0
\end{bmatrix}_A
\xrightarrow{\text{ CG$_0 $}}
\begin{bmatrix}
P_0 & a & 0 & 0 & 0 & b \\
a^\dagger & 0 & 0 & 0 & 0 & c \\
0 & a & 0 & 0 & 0 & 0 \\
0 & 0 & 0 & P_1 & 0 & 0 \\
0 & a & 0 & 0 & 0 & 0 \\
b^\dagger & c^\dagger & 0 & 0 & 0 & P_2
\end{bmatrix}_{SA},
\]
where $a,b$ and $ c $ are the off-diagonal elements of the system's
density matrix. The diagonal elements of the ancilla after tracing out system are $ P_0, P_1+P_2 $.
Thus we can measure $ P_0 $ non-invasively. Similarly for CG$ _1 $ and CG$ _2 $ after the similar procedure, the diagonal elements of
ancilla are $ P_1, P_0+P_2 $ and $ P_2, P_0+P_1 $  respectively, which enables a way of measuring $ P_1 $ and $ P_2 $ non-invasively.
The  measurement at the end of the expriment is not required to be non-invasive since we are not worried about the future dynamcis of the system. After the second evolution of the system, we measure the diagonal elements of the combined ancilla and system state. The elements corresponding to state $ \ket{00}_{SA},\ket{10}_{SA}$, and $\ket{20}_{SA} $ are post-selected. These elements correspond to probabilities, $ P(i,0) $, $ P(i,1) $, and $ P(i,2) $ respectively when CG$ _{i} $ gate is applied, where $ i=0,1,2 $ corresponds to the three states of the system.This scheme is illustrated in Fig. \ref{fig_scheme1}.

Each single measurement described above  follows the L\"uders update rule. However, since we are post-selecting, we end up with only part of the quantum channel (i.e. a subchannel) that corresponds to the negative result.  Adding the three subchannels that we post-select on, effectively creates a measurement that does not follow the L\"uders update rule. Instead the effective trace-preserving channel that describes the evolution during  the measurement is represented by three Kraus operators $\mathbb{K}_i=\ket{i}\bra{i}$.  Nevertheless the measurement is an INRM.

\begin{figure}[htb]
\begin{center}
\includegraphics[width=0.9\columnwidth]{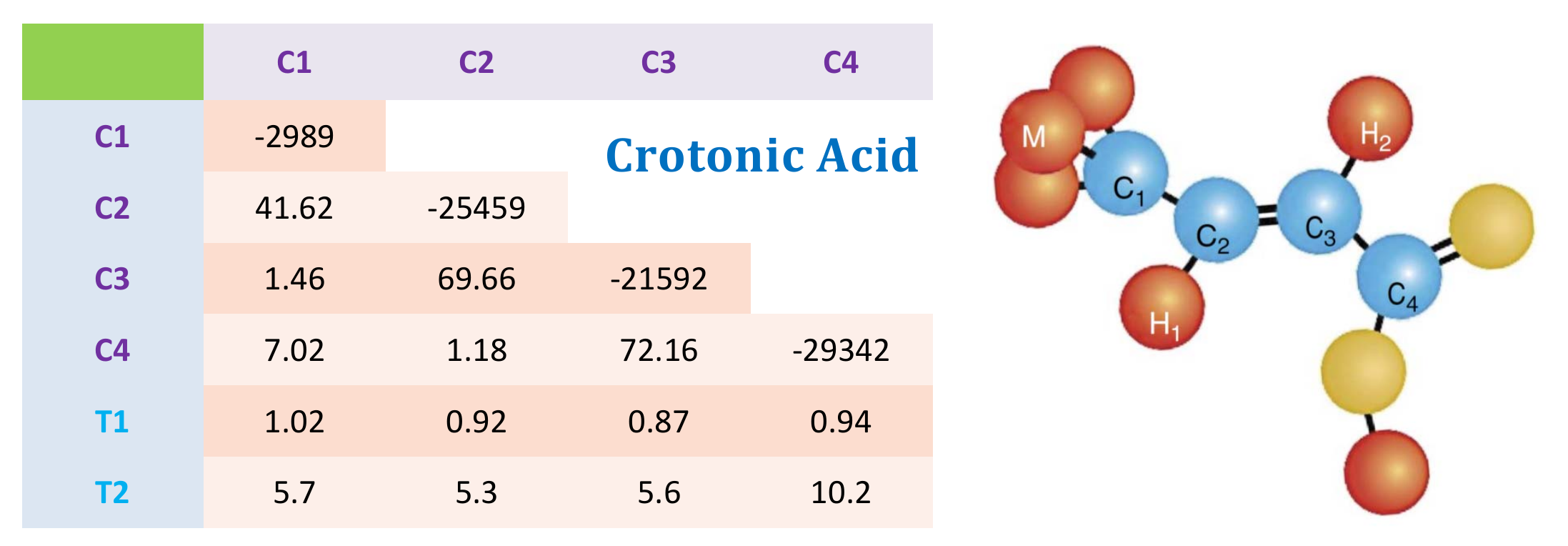}
\end{center}	
	\caption{$^{13}$C-labeled trans-crotonic acid. The table shows the resonance frequencies
		(diagonal elements, in hertz), the J-coupling constants (off-diagonal elements,in hertz), and the relaxation times T$ _1 $ and T$ _2 $ (in seconds). C$_2 $ and C$_3 $ are used to simulate the dynamics of the spin-1 system, and C$_4 $ is used as the ancilla that records the INRMs.}
	\label{fig_molecule}
\end{figure}

\textit{Experimental implementation and results.}---The experiments are carried out at the ambient temperature on a Bruker DRX 700MHz NMR spectrometer.
As described earlier, a spin-1 system and a spin-1/2 ancilla are needed to perform the non-invasive measurements. In the experiments, we use two spin-1/2 nuclei to simulate the dynamics of the spin-1 system via the Clebsch-Gordan approach \cite{shankar2012}, which transforms a space consisting of two spin-1/2 particles to another space consisting of one spin-1 and one spin-0 particle.
This transformation defining the spin-1 in terms of two spin-1/2 particles are
$ \ket{0}_s=\ket{00} $, $ \ket{1}_s=(\ket{01}+\ket{10})/\sqrt{2} $, and $ \ket{2}_s=\ket{11} $, as well as the spin-0 (singlet) state $ \ket{s}=(\ket{01}-\ket{10})/\sqrt{2} $.
For convenience, we employ this spin-1/singlet notation to describe the system state unless otherwise specified.

\begin{figure*}
\begin{center}
\includegraphics[width=1.8\columnwidth]{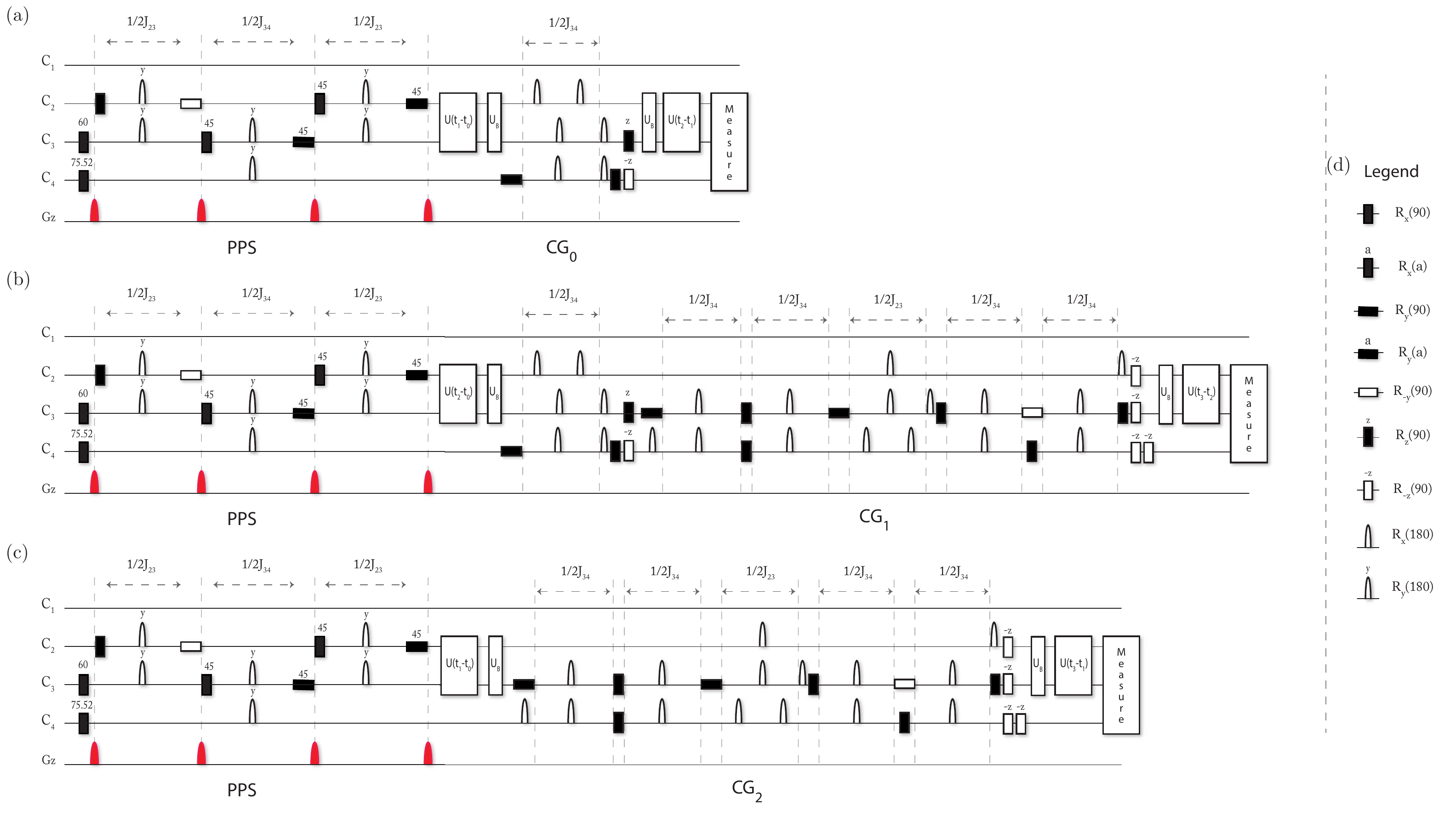}
\end{center}	
	\caption{Pulse sequence for the PPS preparation and controlled gates. Only CG$ _0 $
		and CG$ _2 $ are shown as CG$ _1 $ is just realized by applying CG$ _2 $ after CG$ _0 $. The gate U$_B $ is the Clebsch-Gordan matrix. The first four rows correspond to the pulses acting on different nuclei, and the last row represents the $z$-gradient field.}
	\label{fig_circuit}
\end{figure*}

Therefore, we need three qubits to implement the experiment. The sample is chosen as $^{13}$C-labeled trans-crotonic acid dissolved in acetone-d6.
The molecular structure, Hamiltonian parameters and the relaxation times (T$ _1 $ and T$ _2 $) are shown in Fig. \ref{fig_molecule}, where C$_2 $ and C$_3 $ are used to simulate the dynamics of the spin-1 system and C$_4 $ as  the ancilla.
The spatial averaging method \cite{cory_pps} is adopted to initialize the 3-qubit NMR system into the pseudo-pure state (PPS)
\begin{align}\label{pps}
\rho_{pps}=\frac{1-\epsilon}{8}{\mathbb{I}}+\epsilon\ket{0}\bra{0}_s \otimes\ket{0}\bra{0},
\end{align}
where $\mathbb{I}$ is identity and $\epsilon\approx 10^{-5}$ is the polarization. The NMR circuit of the PPS preparation is shown in Fig. \ref{fig_circuit}(a).

\begin{figure}[htb]
\begin{center}
\includegraphics[width=0.95\columnwidth]{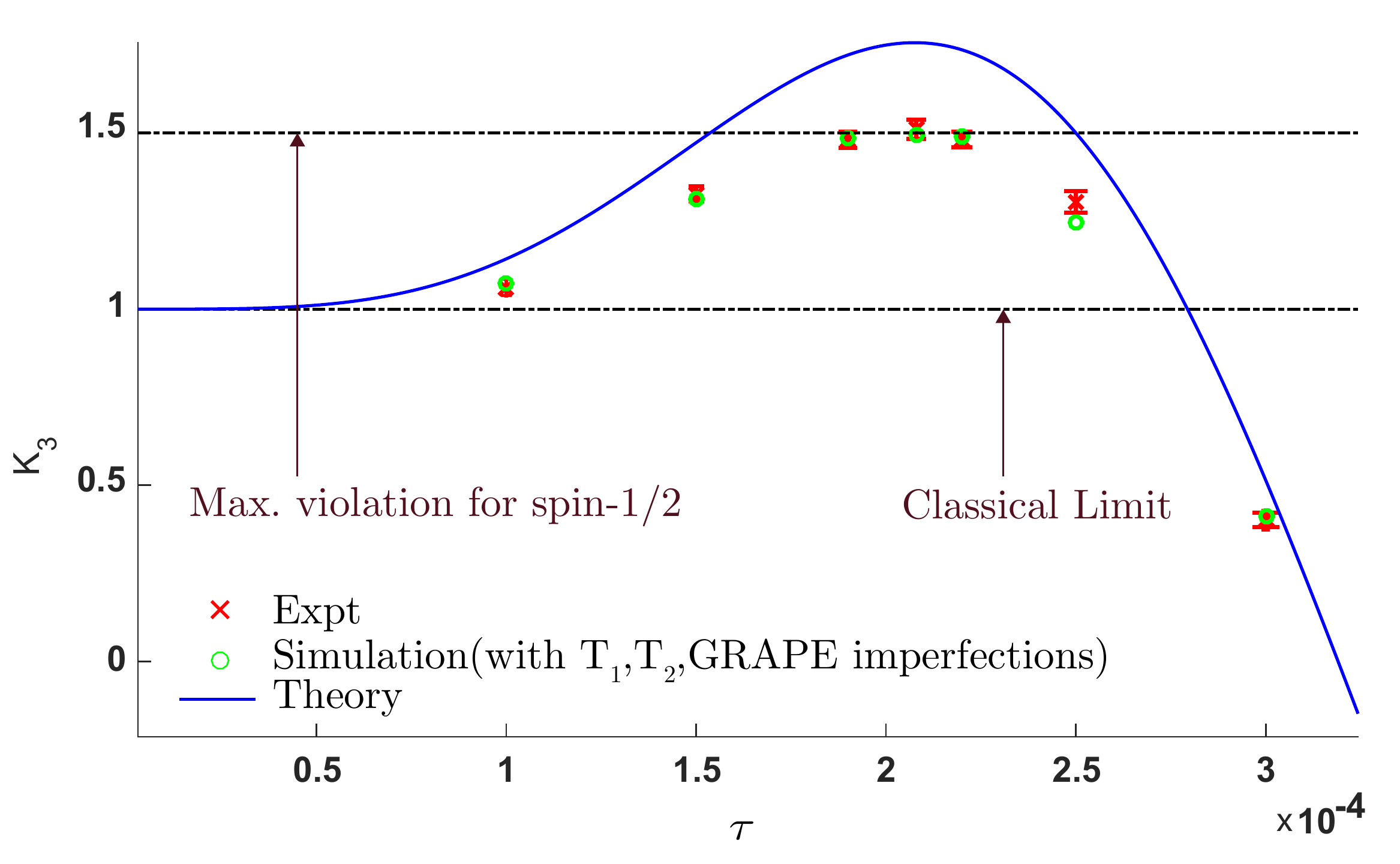}
\end{center}	
	\caption{Experimental values of $ K_3 $ (red crosses) along with the theoretical predictions (blue curve) and simulated results (green circles). $ \tau $ is the tunable time between measurements, i.e. $ \tau = t_2-t_1 = t_3-t_2 $. A violation of the LGI means the value of $K_3$ goes beyond $1$ which is the classical limit. The maximum violation in a 3-level system is $K_3\approx 1.757$ when choosing $\tau=0.208 $ ms. In experiment, decoherence limits our maximum violation around $K_3=1.511 \pm 0.027$. }
	\label{fig_final_resutl}
\end{figure}

The Hamiltonian of the spin-1 system during the free evolutions in Fig. \ref{fig_scheme1} is chosen as $ \mathcal{H}_{sys} = - \Omega \sigma_x^{s_1}/2 $, where $\Omega$ is set as 1 kHz and $\sigma_x^{s_1} $ is the Pauli operator in the spin-1 representation. The propagator at time $t_i$ is thus
\be
U(t_i) = e^{-i2\pi  \mathcal{H}_{sys} t_i}.
\ee
In the experiment, the three different times are chosen as $ t_1 = 0.5$ ms, $t_2 = \tau + t_1$, and $t_3 = \tau + t_2$ respectively, and the experiments are conducted for a few values of $\tau$ as shown in Fig. \ref{fig_final_resutl}. The observable to be measured is chosen as $ Q = -\ket{0}\bra{0}_s+\ket{1}\bra{1}_s+\ket{2}\bra{2}_s$, which is equivalent to measuring the diagonal elements of the density matrix. Ideally, the maximal value of $K_3$ should be obtained at $ \tau=0.208 $ ms, and the following tests of non-invasiveness are performed at this optimal point.

The controlled gates in Fig. \ref{fig_scheme1} are decomposed into single-qubit rotations and delays, and the pulse sequence of the entire experiment is illustrated in Fig. \ref{fig_circuit}.
All pulses are realized by the gradient ascent pulse engineering (GRAPE) technique \cite{fortunato,maheshsmp,khaneja}, and are robust against the B$_1$ inhomogeneity with the fidelity over
0.997. The $\pi/2$ and $\pi$ pulses are of length 1 ms.
The observable $Q$ is measured by performing  diagonal tomography in the spin-1 subspace without considering the spin-0
component \cite{tomopra}.

The values of $K_3$ for different $\tau$ are shown in Fig. \ref{fig_final_resutl}, where the blue curve is the theoretical prediction, green circles are the simulated results with the T$_1$, T$_2$ and pulse imperfections incorporated, and red crosses are the experimental results.
At the point of the maximum violation, $\tau=0.208 $ ms, the experimental values of correlations are $ \langle Q_1 Q_2 \rangle =0.542 \pm 0.021 $, $  \langle Q_2 Q_3 \rangle = 0.294 \pm 0.016 $, and $ \langle Q_1 Q_3 \rangle =-0.676 \pm 0.003 $, respectively. It leads to the experimental value of $K_3=1.511 \pm 0.027$, in consistence with the simulated result 1.495. In contrast, the ideal value of the maximum violation is 1.757, and the discrepancy ($\approx 0.246$) between the experimental and ideal value is dominated by the T$_1$, T$_2$ relaxation, as the pulse imperfections merely contribute around 0.01 loss of the ideal value.

\textit{Experimental test of assumptions.}---In getting the theoretical bound of $K_3$ we have made a number of implicit assumptions about our experimental system, in which the most notable assumption is INRMs. Since it is possible to modify the LGI by taking into account any deviations from these assumptions, our experiment is supplemented by another set of experiments to test (i) the invasiveness of the intermediate measurements and loss, (ii) preparation errors, and (iii) malicious losses. We also discuss the possibility of dark counts. An additional assumption about the pseudo-pure state is discussed in the appendix.

First we quantify how much the system is disturbed due to the imperfect controlled gates. Ideally these controlled gates should not disturb the system when it is in a fixed state $\ket{0}_s,\ket{1}_s,\ket{2}_s$ or $\ket{s}$, but  in practice they do disturb the system due to the long application time and pulse imperfections. Moreover, the three controlled gates are distinct and are expected to have different back actions on the system even after the negative results are post-selected in the INRMs.  Explicitly, CG$_0 $ is a direct J-coupling gate,  CG$_2 $ involves two SWAP gates and CG$_1 $ is a combination of the two. The experimental lengths of CG$_0 $, CG$_1 $ and CG$_2 $ are about $ 40 $ ms, $ 116 $ ms and $ 76 $ ms, respectively.
In attempt to quantify how much the system is disturbed by INRMs, we perform the following two types of experiments:
(a) start with either $ \ket{0}_s $, $ \ket{1}_s $, or $ \ket{2}_s $, evolve the system for a fixed time and measure the probabilities;
(b) start with either $ \ket{0}_s $, $ \ket{1}_s $, or $ \ket{2}_s $, apply the controlled gate and measure the probabilities.
Ideally, the results from the two experiments should match perfectly, but they indeed have variations in the presence of errors.
Table \ref{table:Table2} shows the experimental results and their contribution to the inequality is discussed in next section.

 In testing non-invasiveness we can calculate the correlation, $ C $,  value when the starting state is $ \ket{p}_s $($ p=0,1,2 $) using eq. \eqref{eq:correlation})
\begin{align}
C_{\ket{p=0}}&= P(0,0)-P(0,1)-P(0,2) \nonumber\\
C_{\ket{p=1,2}}&= -P(p,0)+P(p,1)+P(p,2)
\label{eq:Qforloss}
\end{align}

Now the difference between the $ C $ value when we apply the gate vs no gate is the disturbance induced by our measurements.
This $ \Delta C $ values contribute thrice in calculation of LGI (eq.\ref{eq:k3}).
Since it contributes 2 times positively and one time negatively, we define the following modification over the original inequality
\begin{equation}
KM1 = -\min(\Delta C \pm P.E.,0)+2\max(\Delta C \pm P.E.,0)
\label{eq:KM1}
\end{equation}
where $ P.E. $ is the preparation error, i.e. how much the initial state deviates from the expected.

The values of probabilities which were not used in eq. \ref{eq:Qforloss} are considered as loses. We consider the losses that can act maliciously during the experiment, i.e we assume that the losses are somehow maliciously designed to maximize $K_3$. The lowest value when we apply no gate is considered non malicious since it is independent of the gate and/or initial state.  The difference between the highest and lowest give the range for the possible malicious errors. Now this is the value for one evolution, in the LG experiment there are 5 such evolutions (two for each of the experiments giving $Q_1Q_3$ and $Q_2Q_3$ and one for the experiment giving $Q_1Q_2$), hence the total malicious loss if 5 times the difference

With these modifications, we modify the original inequality on  $ K_3 $ to
\begin{equation}
K_3 \leq 1+KM1+Mal = 1+0.1936+0.2095=1.4031
\end{equation}

We note that while this value takes the imperfections in preparation into account in the worst possible way, it is extremely unlikely that these preparation errors decrease the discrepancy between the ideal measurements and the actual measurements. A slightly more liberal version of the inequality would read

\begin{eqnarray}
&& KM1 = -\min(\Delta C  ,0)+2\max(\Delta C ,0) \\
&& K_3 \leq 1+KM1+Mal = 1+0.0912+0.2095=1.3007 \nonumber
\end{eqnarray}

Finally, we must account for the sources of errors that lead to \emph{dark counts}, i.e an artificial increase in the probabilities that are post selected. There are two possible sources for this kind of error. First the measurements are not perfect and there are situations where the ancilla does not rotate to $\ket{1}$ when it should, leading to a false reading of $\ket{0}$. Second, there are situations where a system in the singlet state goes back into one of the triplet states. The margin for the violation leaves us with an upper bound on the tolerance of the violation for possible dark counts, assuming these behave in the most malicious way possible. These can range between $ 0.1081 $ and $ 0.2105 $ depending on how we account for preparation errors in the test of non-invasiveness.

\textit{Discussion.}---The motivation behind a LG experiment is to test macroscopic realism, i.e try to refute  MR  for a \emph{macroscopic} system or at least convince a skeptic that MR assumption  is implausible. While the NMR sample that we use can be considered macroscopic, the individual molecules are still in the microscopic domain, moreover there is little doubt that the individual nuclear spins can be in a superposition state. In that respect it is not too surprising that the LGI is violated, and indeed its violation tells us nothing  new about macroscopic realism. We do, however, learn that we can control the systems well enough to violate the inequality and that the qutrit used   can pass some quantum tests under reasonable assumptions. The violation of a LGI does not rule out the existence of a hidden variable model and indeed a skeptic could simply argue that our system behaves strangely due to the existence of hidden variables that are influenced by our choice of measurements.   For liquid state NMR we already know that such a model exists \cite{Menicucci2002}.  Moreover we purposely discarded some of the experimental data as part of the experiment, i.e the spectrum generated at the end of each experiment could be used to generate more than the six probabilities we discussed (the off diagonal elements in the density matrix).

Since we are not, strictly speaking, testing MR,  our main result is not the violation  per-se but rather the methods used to achieve the violation, the discussion of possible errors in the experiment and the demonstration of their experimental relevance.  Such a discussion has been missing from much of the experimental literature to date (see \cite{Knee2016,knee2012violation} for two exceptions).  The LG test cannot be performed without some assumptions about the physical systems involved and, in particular, the inner workings of the measurements that we assume are non invasive. These assumptions must be tested, as they can lead to artificial violations of the inequality. In our experiment we tested particular malicious scenarios that, although unlikely, must be taken into account and discussed before they are rejected (experimentally or theoretically).  We note that both our simulated  predictions and experimental results (see Fig. \ref{fig_final_resutl}) show that such artificial violations are unlikely in our system, consequently we believe that although many previous experiments did not include a careful analyses of possible errors, the violations of LGI in those experiments would probably hold even if imperfections were taken into account.

\textit{Conclusion.}---We demonstrated a violation of 3-level LGI. Non-invasive measurements, an  essential requirement when performing a LG test were carried out  using  ideal negative result measurement. We verified  the non-invasiveness of such measurements experimentally and quantified  how much this measurement disturbs the system. We also took  account  error that can occur in experiments into account and used them to modify the original inequality. These modifications resulted  in increasing the classical bound and making our violation significantly smaller (but still beyond the error margins). We emphasize that  the margin of violation between quantum and classical upper bound is greater when a 3-level system is tested (compared to  a 2-level system). In practice the actual margin is quite low when various errors are taken in account and the use of the modified (non L\"uders) measurement scheme allowed us to observe the violation despite many imperfections.  The difference in experimental value from theoretical is due to the T1 and T2 decay, these errors can be avoided in different systems, for example if the couplings are strong, the gate lengths will be short.  It would be a challenge to the quantum control community to observe a violation larger than $0.5$ above the classical bound (modified for imperfections), however the real challenge remains to demonstrate such violations in macroscopic systems.

\textit{Acknowledgments.}---We thank A. Leggett for discussions and suggestions for improving the experiment and testing non-invasiveness, Annie Park, Daniel Park and Guanru Feng for helpful discussions and comments,
This work is supported by Industry Canada, NSERC and CIFAR. AB is now at the University of Toronto. 

\bibliographystyle{apsrev4-1}
\bibliography{bib_ch}

\clearpage
\onecolumngrid
\appendix

\section*{Appendix}

\subsection*{Experimental  data for test of the measurement procedure }

\begin{table}[htb]
	\centering
	\begin{tabular}{|c||c|c|c||c|c|c||c|c|c||}
		\hline
		& \multicolumn{3}{c||}{Measure at time, t$_1$ and t$_2$} & \multicolumn{3}{c||}{Measure at time, t$_2$ and t$_3$} & \multicolumn{3}{c||}{Measure at time, t$_1$ and t$_3$}\\
		\hline
		& Theory & Sim & Exp & Theory & Sim &  Exp & Theory & Sim & Exp \\
		\hline
		\hline
		$00$ & 0.0000 & 0.0003 & 0.0317 $ \pm $ 0.0134 & 0.0778 & 0.0765 & 0.0949 $ \pm $ 0.0025 & 0.0000 & 0.0306 & 0.0289 $ \pm $ 0.0033 \\
		\hline
		$01$ & 0.0000 & 0.0333 & 0.0703 $ \pm $ 0.0054 & 0.0636 & 0.0758 & 0.0497 $ \pm $ 0.0028 & 0.0000 & 0.0007 & 0.0486 $ \pm $ 0.0018 \\
		\hline
		$02$ & 0.0000 & 0.0131 & 0.0020 $ \pm $ 0.0029 & 0.0186 & 0.0021 & 0.0090 $ \pm $ 0.0001 & 0.0000 & 0.0001 & 0.0033 $ \pm $ 0.0020 \\
		\hline
		$10$ & 0.0000 & 0.0174 & 0.0142 $ \pm $ 0.0027 & 0.2170 & 0.2036 & 0.1876 $ \pm $ 0.0047 & 0.0000 & 0.0467 & 0.0090 $ \pm $ 0.0041 \\
		\hline
		$11$ & 0.0000 & 0.0253 & 0.0755 $ \pm $ 0.0038 & 0.0318 & 0.0183 & 0.0325 $ \pm $ 0.0079 & 0.0000 & 0.0030 & 0.0410 $ \pm $ 0.0022 \\
		\hline
		$12$ & 0.0000 & 0.0195 & 0.0027 $ \pm $ 0.0010 & 0.2170 & 0.2294 & 0.2331 $ \pm $ 0.0043 & 0.0000 & 0.0048 & 0.0009 $ \pm $ 0.0006 \\
		\hline
		$20$ & 0.1364 & 0.1487 & 0.1680 $ \pm $ 0.0005 & 0.0542 & 0.0781 & 0.1175 $ \pm $ 0.0112 & 0.8682 & 0.7771 & 0.7925 $ \pm $ 0.0054 \\
		\hline
		$21$ & 0.4659 & 0.4061 & 0.3616 $ \pm $ 0.0026 & 0.1853 & 0.1681 & 0.1713 $ \pm $ 0.0024 & 0.1272 & 0.1372 & 0.1064 $ \pm $ 0.0013 \\
		\hline
		$22$ & 0.3977 & 0.3378 & 0.3245 $ \pm $ 0.0010 & 0.1582 & 0.1497 & 0.1261 $ \pm $ 0.0038 & 0.0046 & 0.0144 & 0.0004 $ \pm $ 0.0002 \\
		\hline
		sum  & 1	& 1.0015 & 1.0504 & 1 & 1.0029 &  1.0217& 1 &  1.0146  & 1.0309 \\
		\hline
	\end{tabular}
	\caption{{\bf Experimental results for the setting that leads to a maximal LG violation.} Each table shows the result for a single setting (see Fig. \ref{fig_lgi}). The row index denotes the two measurement outcomes and the three values (Theory, Sim, Exp) correspond to the probabilities for these outcomes in theory, simulation and experiment respectively. (For example the row $01$ represents the probability that the result was 0 in the first measurement and 1 in  the second).  Since the results are post-selected, the probabilities in the simulation and experiment do not add up to 1. }
	\label{table:Table1}
\end{table}

		\begin{table}[htb]
			\centering
			\begin{tabular}{|c||c|c||c|c||c|c||}
				\hline
				& \multicolumn{2}{c||}{Starting state = $\outpr{0}{0}$} & \multicolumn{2}{c||}{Starting state =$ \outpr{1}{1}$} & \multicolumn{2}{c||}{Starting state =$ \outpr{2}{2}$}\\
				\hline
				& $NG$ & CG$_0$ & $NG$ & CG$_1$ & $NG$ & CG$_2$ \\
				\hline
				\hline
				 \cellcolor{blue!25}$00$ & 0.3885$\pm$0.0022 & 0.3582$\pm$0.0021 & 0.4297$\pm$0.0017 & 0.4133$\pm$0.0034 & 0.1343$\pm$0.0012 & 0.1570$\pm$0.0015 \\
				\hline
				 \cellcolor{red!25}$01$ & 0.0001$\pm$0.0002 & 0.0323$\pm$0.0040 & 0.0071$\pm$0.0012 & 0.0273$\pm$0.0016 & 0.0001$\pm$0.0003 & 0.0164$\pm$0.0023 \\
				\hline
				 \cellcolor{blue!25}$10$ & 0.4143$\pm$0.0055 & 0.3974$\pm$0.0023 & 0.0521$\pm$0.0023 & 0.0570$\pm$0.0013 & 0.4023$\pm$0.0028 & 0.3637$\pm$0.0026 \\
				\hline
				 \cellcolor{red!25}$11$ & 0.0006$\pm$0.0007 & 0.0147$\pm$0.0012 & 0.0091$\pm$0.0032 & 0.0293$\pm$0.0019 & 0.0290$\pm$0.0022 & 0.0472$\pm$0.0018 \\
				\hline
				 \cellcolor{red!25}$S0$ & 0.0525$\pm$0.0026 & 0.0370$\pm$0.0067 & 0.0745$\pm$0.0022 & 0.0672$\pm$0.0028 & 0.0656$\pm$0.0026 & 0.0721$\pm$0.0014 \\
				\hline
				 \cellcolor{red!25}$S1$ & 0.0003$\pm$0.0002 & 0.0011$\pm$0.0018 & 0.0040$\pm$0.0002 & 0.0021$\pm$0.0001 & 0.0004$\pm$0.0002 & 0.0002$\pm$0.0002 \\
				\hline
				 \cellcolor{blue!25}$20$ & 0.1428$\pm$0.0025 & 0.1351$\pm$0.0029 & 0.4219$\pm$0.0019 & 0.4031$\pm$0.0031 & 0.3680$\pm$0.0017 & 0.3431$\pm$0.0012 \\
				\hline
				 \cellcolor{red!25}$21$ & 0.0009$\pm$0.0003 & 0.0242$\pm$0.0021 & 0.0016$\pm$0.0002 & 0.0007$\pm$0.0003 & 0.0003$\pm$0.0006 & 0.0002$\pm$0.0001 \\
				\hline
	\end{tabular}
	\caption{\textbf{Experimental results for the test on non-invasiveness.} Each  of the three tables  shows the result when starting with the state mentioned on the top. The rows corresponds to the probabilities of the state denoted in first column. The first and second index in first column corresponds to the state of system and ancilla respectively. CG$_i$ stands for the gate applied and $ NG $, when no gate is applied. Ideally the column $NG$ should contain positive values only for the states $\ket{00},\ket{10}$ and $\ket{20}$ (in blue), all other values are treated as losses since they are lost in post-selection.  Moreover, for  an INRM the columns NG and $CG_i$ should match, the discrepancies between these columns at the post selected values (blue) are used to give an upper bound on the possible deviation from $K_3$ due to the measurement procedure.}
	\label{table:Table2}
\end{table}

\newpage
\begin{table}[htb]
	\centering
	\begin{tabular}{c||cc||cc||cc}
		\hline
		& \multicolumn{2}{c||}{Starting state = $\outpr{0}{0}$} & \multicolumn{2}{c||}{Starting state = $\outpr{1}{1}$} & \multicolumn{2}{c}{Starting state = $\outpr{2}{2}$}\\
		\hline
		& $NG$ & CG$_0$ & $NG$ & CG$_1$ & $NG$ & CG$_2$ \\
		\hline
		\hline
		\hline
		$ Q  $   & -0.1686  & -0.1743  &  0.0443  & 0.0468  &  0.6360  &  0.5498 \\
		\hline
		$ \Delta Q= Q_G-Q_{NG}$& \multicolumn{2}{c||}{$-0.0057$} & \multicolumn{2}{c||}{$0.0025$} & \multicolumn{2}{c}{$ -0.0862$} \\
		\hline
		P.E.& \multicolumn{2}{c||}{$0.0187$} & \multicolumn{2}{c||}{$ 0.0436$} & \multicolumn{2}{c}{$0.0152$} \\
		\hline
		$ \Delta Q- $P.E.& \multicolumn{2}{c||}{$-0.0244$} & \multicolumn{2}{c||}{$-0.0411$} & \multicolumn{2}{c}{$-0.1014$} \\
		\hline
		$ \Delta Q+$P.E.& \multicolumn{2}{c||}{$0.0130$} & \multicolumn{2}{c||}{$0.0461$} & \multicolumn{2}{c}{$-0.0710$} \\
		\hline			
		$ KM1 $ & \multicolumn{6}{c}{$0.1936$} \\
		\hline
		$ Non. Mal $ & \multicolumn{6}{c}{$0.0544$} \\ 						
		\hline
		$ Mal $ & \multicolumn{6}{c}{$0.0419*5=0.2095$} \\ 						
		\hline
	\end{tabular}
	\caption{\textbf{ Summary of imperfections calculated in the test of non-invasiveness.} The table list various modification made to the original LG as explained in text. Loss is  calculated using the discarded values  in an experiment where we don't expect to discard any values in post-selection (the red columns in table \ref{table:Table2} are discarded in post selection, and the loss is the sum of these values). $ Q $ is calculated using equation-(\ref{eq:Qforloss}), $\Delta Q $ is the difference of $ Q $ values when the gate is applied and when it is not. For an INRM and the setup used $\Delta Q=0$ and any deviation from $0$ could theoretically boost $K_3$ even in a MR system.  P.E. stands for preparation error, i.e  the  probability that the prepared starting state is not the desired starting state. KM1 is the maximal boost to $K_3$ due to measurement error, as defined in equation-(\ref{eq:KM1}). The losses are broken into two types.  Non-Malicious (Non. Mal.) are the losses that appear irrespective of the specific experiment.  Malicious ( Mal)   are the losses that may depend on the choice of experiment. We assume the malicious losses are chosen in such a way that they boost the calculated value $K_3$  by as much as possible. }
	\label{table:final}
\end{table}

\subsection*{The  pseudo pure state dynamics.} 
\begin{figure}[htb]
	\hspace*{-2cm}
	\centering
	\includegraphics[scale=.8]{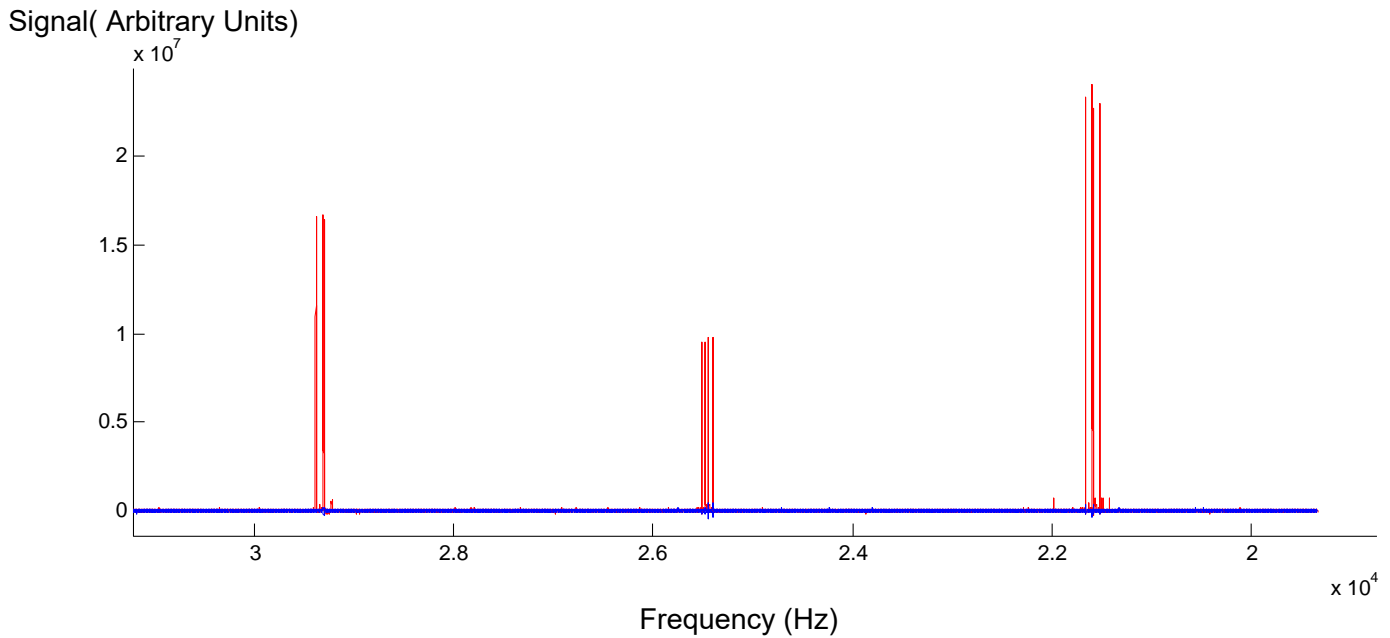}
	\vspace{-9cm}	
	\caption{{\bf the spectra for the LG experiment with the identity as the initial state}.  The blue spectra is the signal for a run of the Leggett-Garg  experiment  with the identity as the initial state.  The red spectra is the initial thermal state which is given as a reference. Note that while an identity will give a flat spectrum at 0, the flat spectrum does not guarantee that the state is the identity. To verify that this is the identity we rotated the state before the final measurement and produced the same flat spectrum for different observables.  
	\label{fig_final_id}}
\end{figure}

In an NMR experiment we have access to pseudo-pure states(Eq, 6 Main text), to verify that this does not effect the credibility of the result, 
we perform the Leggett-Garg experiment starting from an identity state instead of $ \ket{0} $ for the system. 
If starting from an identity state the end state remains identity it has no contribution in the Leggett-Garg inequality.
The spectra for the Leggett Garg test on the identity was compared with a reference spectra of an initial thermal state  (see fig. \ref{fig_final_id}) to ensure that the contribution of the signal is below the level of precision used in the experiment. 

\end{document}